\documentstyle[epsfig,longtable]{aipproc}
\newcommand{\ccbar}{c\overline{c}}

\newcommand{\picb}{\mbox{pb}^{-1}}
\newcommand{\gev}{GeV}
\newcommand{\mev}{MeV}
\newcommand{\dstar}{D^*}
\newcommand{\dstarpm}{D^{*\pm}}
\newcommand{\dsubs}{D_s}
\newcommand{\dsubspm}{D_s^{\pm}}
\newcommand{\dplus}{D^+}
\newcommand{\donen}{D_1(2420)}
\newcommand{\dtwon}{D_2^{*}(2460)}
\newcommand{\jpsi}{J/\psi}
\newcommand{\qsq}{Q^2}
\newcommand{\dstarplus}{D^{*+}}
\newcommand{\dstprplus}{D^{*'+}}
\def\d0{D^{0}}
\def\dsubsfull{ {\dsubs}^{\pm} \rightarrow \phi \pi^{\pm} \rightarrow K^{+}K^{-}\pi^{\pm}}
\def\sgepdsubs{ $\sigma_{ep \rightarrow D_s X}$}
\def\xsecra{0.41\pm0.07^{+0.03}_{-0.05}}
\def\xsepdstar{ \sigma_{ep \rightarrow D^* X}}
\def\xsepdsubs{ \sigma_{ep \rightarrow D_s X}}
\def\kpipis {M(K\pi\pi_s)}
\def\kpiext {M(K\pi\pi_s\pi_4)}
\def\kpiexte {M(K\pi\pi_s\pi_4\pi_5)}
\begin{document}
\title{Recent Results on Charm Photoproduction}

\author{Don Hochman$^*$    
\thanks {Talk given at the PHOTON 2000 Conference, Ambelside, UK,
August 26-31, 2000, to appear in the Proceedings.}
\\
  On behalf of the ZEUS Collaboration}
\address{$^*$Weizmann Institute of Science,Rehovot Israel 76100}          

\maketitle

\begin{abstract}
Photoproduction of $D_s^{\pm}$ mesons has been measured in the ZEUS detector at
HERA and compared with predictions of NLO pQCD calculations.  The ratio of 
$D_s^{*\pm}$ to $D^{*\pm}$ cross sections has been compared to results from
$e^{+}e^{-}$ experiments.  Orbitally excited P-wave charm
mesons have been observed in the $D^{*\pm}\pi^{\mp}$ final state.
The fraction of $D^{*\pm}$ 's originating from these mesons has
been calculated and compared with that from $e^{+}e^{-}$ interactions.
No evidence for radially excited mesons decaying to $D^{*\pm}\pi^{+}\pi^{-}$
was found.         
The inelastic production of J/$\psi$ mesons has been
measured and compared to LO and NLO pQCD predictions.
\end{abstract}

\section*{Introduction}
 
Charm photoproduction measurements have been performed at the 
HERA {\it ep} collider in the ZEUS detector from data taken during 1995-2000.  
Electrons or positrons with energy $E_e = 27.5 \, \gev$
collided with protons of energy $E_p = 820 \, \gev$ \, (1995-1997) or
$E_p = 920 \, \gev$ \, (1998-2000).
The ZEUS detector description can be found elsewhere 
\cite{detectr}.
 
The  decay chain $\dsubsfull$ ($38 \picb$ integrated luminosity) was studied 
\cite{dsubsrf} 
as a continuation of a previous analysis of charm photoproduction 
\cite{dstarrf}.
The study of $\dsubspm$ photoproduction provides another test of next-to-leading
order (NLO) perturbative quantum chromodynamics (pQCD) calculations.
 
Orbitally excited P-wave $D$ mesons can decay to a $\dstar$ by
pion emission.  Two of these states ($\donen$ and $\dtwon$)
have been found to decay into narrow states 
\cite{pdg} 
with properties predicted 
by Heavy Quark Effective Theory (HQET) 
\cite{hqet} 
and a third broad state has been seen by the CLEO collaboration
\cite{cleo1}.
A radial excitation of the $\dstarpm$
with a mass of about 2.6 \, $\gev$ decaying into   
$D^{*\pm}\pi^{+}\pi^{-}$  
has been reported by DELPHI 
\cite{delphi}
but not seen by OPAL and CLEO
\cite{opcleo,rodr}.
 
Inelastic $\jpsi$ photoproduction proceeds via direct (resolved) processes, 
where the virtual photon (parton from the photon) interacts with a
parton from the incoming proton.  In the dominant process,
boson gluon fusion (BGF), the latter parton is a gluon.
Photon diffraction to $\jpsi$ also contributes.  The inelasticity variable, 
$z = \frac{P\cdot p_\psi}{P\cdot q}$, 
can be used to distinguish these processes.
Here $P$, $p_\psi$ and $q$ are the four-momenta of the incoming proton, $\jpsi$ 
and exchanged photon, respectively. 
From previous ZEUS data
\cite{psizeus}
the diffractive process dominates at $z > 0.9$, the direct photon process    
dominates at $0.4 < \, $z$ \, < 0.9$.  The resolved photon contribution is
expected to dominate at $z \lesssim 0.2$
\cite{psiresolv}.
 
Color singlet and color octet models have been used to calculate
the above non-diffractive production processes in pQCD.  For the
former, the charm-anticharm pair ($\ccbar$) from the hard process is
identified with the physical $\jpsi$ state.  In this model in leading order (LO)
only the BGF diagram contributes to the direct channel.  In the color octet model
the $\ccbar$ pair from the hard process emits one or more soft gluons
to evolve into the physical $\jpsi$ state.  The free parameters of the
model can be extracted from $\jpsi$ cross-section measurements and
used in other inelastic $\jpsi$ production experiments.
 
\section*{$D_s^{\pm}$ Photoproduction}
$\dsubspm$ production was studied for:
$\qsq \, < \, 1.0 \, \gev^2$,$130 \, < \, W_{\gamma p} \, < \, 280 \, 
\gev$,$3 \, < \, p_{\perp}^{\dsubs} \, < \, 12 \, \gev$,
$|\eta^{\dsubs}| \, < \, 1.5$, where $\qsq$ is the photon virtuality, 
$W_{\gamma p}$ is the virtual photon proton center of mass energy, 
$p_{\perp}^{\dsubs}$ is the transverse momentum of the $\dsubspm$ and 
$\eta^{\dsubs}$ is the  pseudorapidity of the $\dsubspm$.  The effective
mass of two opposite charge track combinations, assumed to be kaons, was
calculated and plotted in Fig. 1a.  A clear enhancement at the $\phi$
mass is seen.  The effective mass of the combinations in this enhancement
region and another track assumed to be a pion was then obtained.  The peak
in the $\dsubspm$ mass region contained $339 \pm 48 \, \dsubspm$ mesons (Fig. 1b),
corresponding to a cross section of
\sgepdsubs $= 3.79 \pm 0.59(stat) ^{+0.26}_{-0.46}(syst) \pm 0.94(br) \, nb$.
\begin{figure}[h!] 
\unitlength1cm
\hspace*{2.5cm}
{\epsfig{file=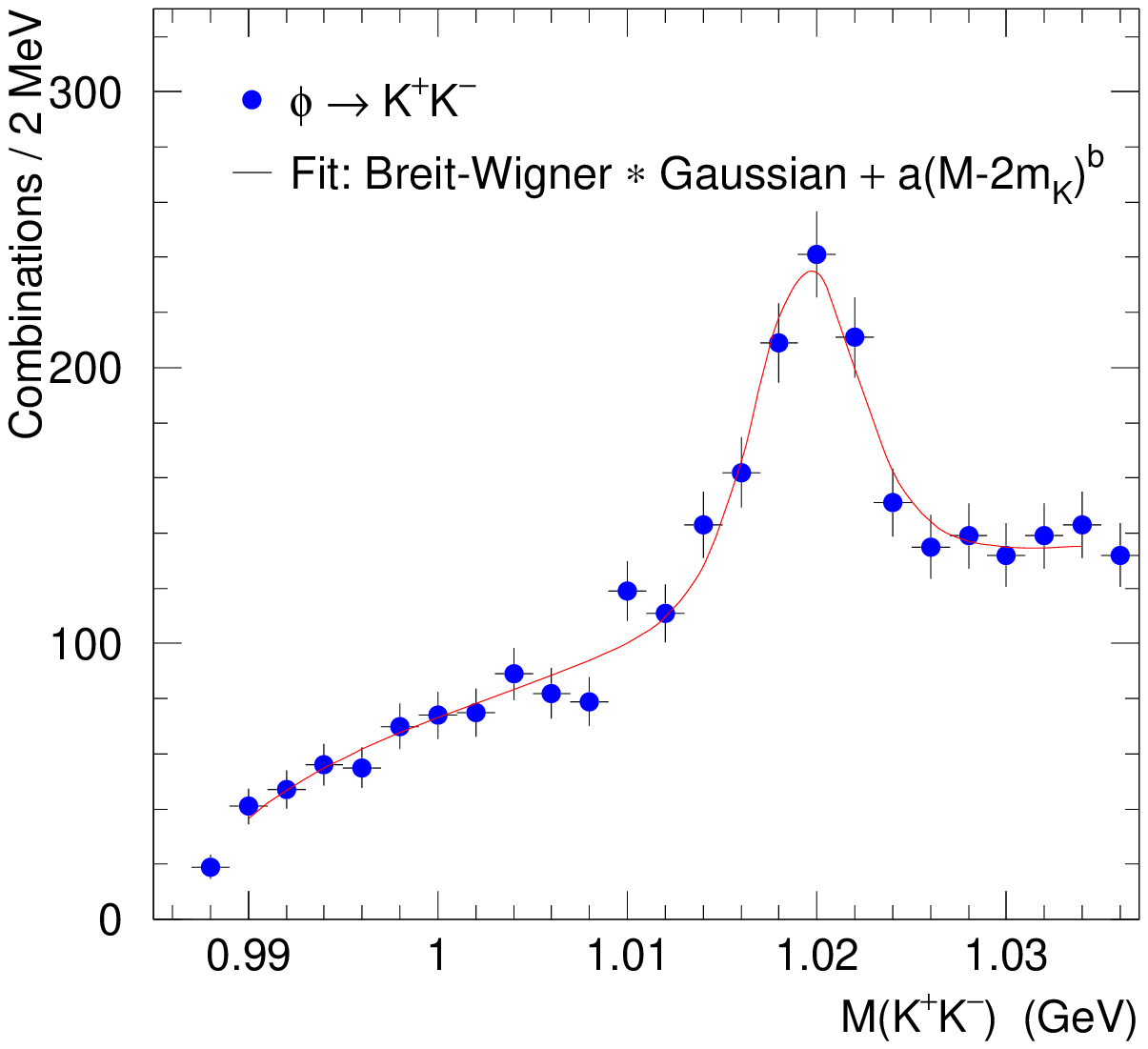,height=5.0cm,width=5.0cm}}
 
\vspace*{-5.0cm}
\hspace*{8.0cm}
{\epsfig{file=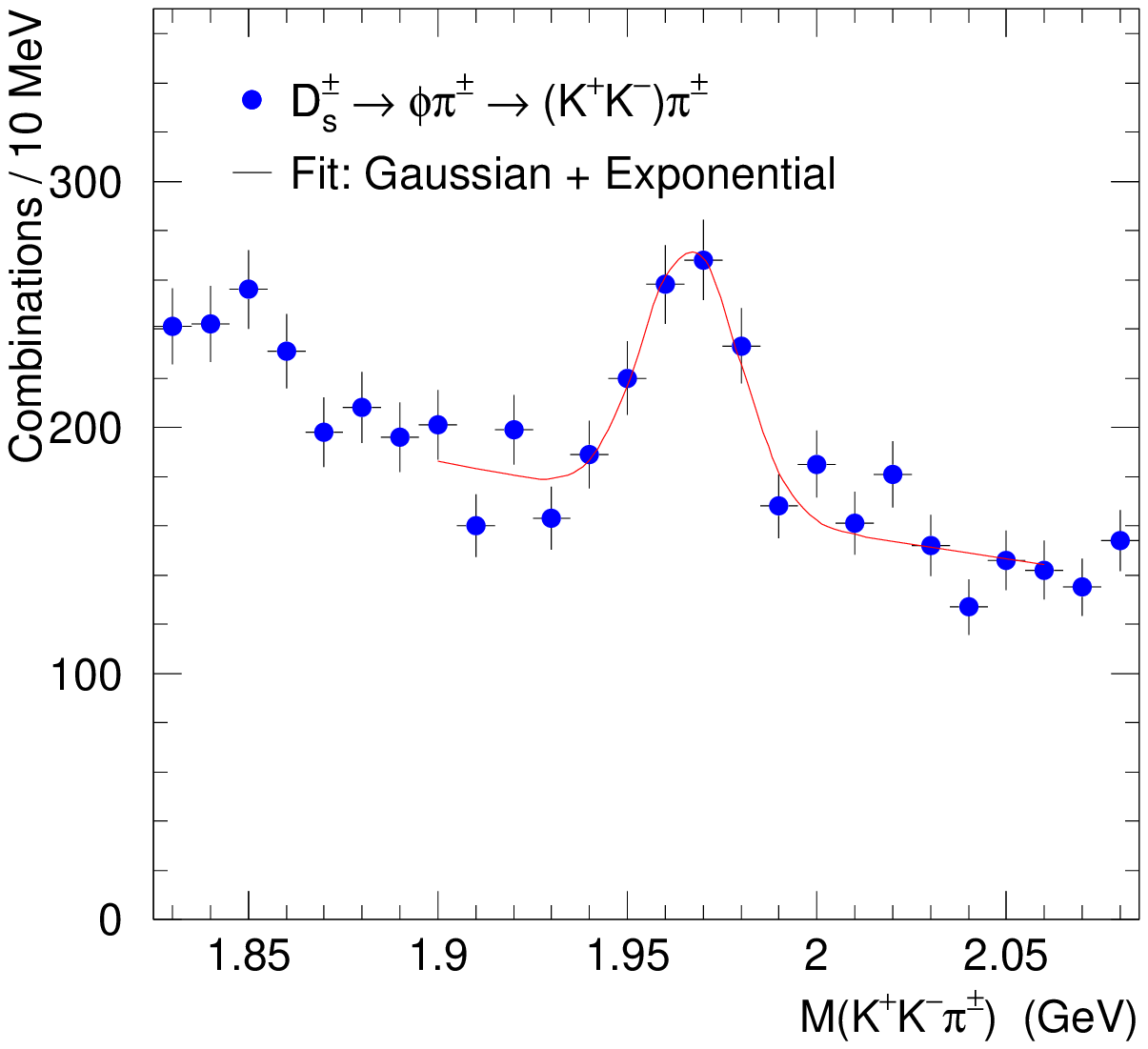,height=5.0cm,width=5.0cm}}
 
\vspace*{-0.1cm}
\bf {~~~~~~~~~~~~~~~~~~~~~~~~~~~~~~~~~(a)~~~~~~~~~~~~~~~~~~~~~~~~~~~~~~~(b)}
\vspace{10pt}   
\caption[]{(a) $M(K^{+}K^{-})$ distribution for events inside the $\dsubspm$ 
mass range,  
$(1.94  <  M(K^{+}K^{-}\pi^{\pm})  <  2.00 \, \gev)$.
The solid curve is a fit to a Breit-Wigner convoluted with 
a Gaussian-shaped resonance and a background parameterization, 
$a[M(K^{+}K^{-}) - 2m_K]^b$. (b) 
$M(K^{+}K^{-}\pi^{\pm})$ distribution for events in the 
$\phi$ mass range,  
($1.0115 < M(K^{+}K^{-}) < 1.0275 \, \gev)$. 
The solid curve is a fit to a Gaussian plus  
an exponential background.}
\label{fig1}
\end{figure}
 
Distributions in $p_{\perp}^{\dsubs}$ and $\eta^{\dsubs}$ were
compared with those for $\dstarpm$ production 
\cite{dstarrf}
and with a fixed order NLO calculation 
\cite{frix} 
in which charm was produced by the BGF process.
The signal is above the prediction (Fig. 2), particularly 
for $\eta$ along the proton beam direction, as was the case for $\dstarpm$
production.
 
\begin{figure}[h!] 
\unitlength1cm
\hspace*{2.5cm}
{\epsfig{file=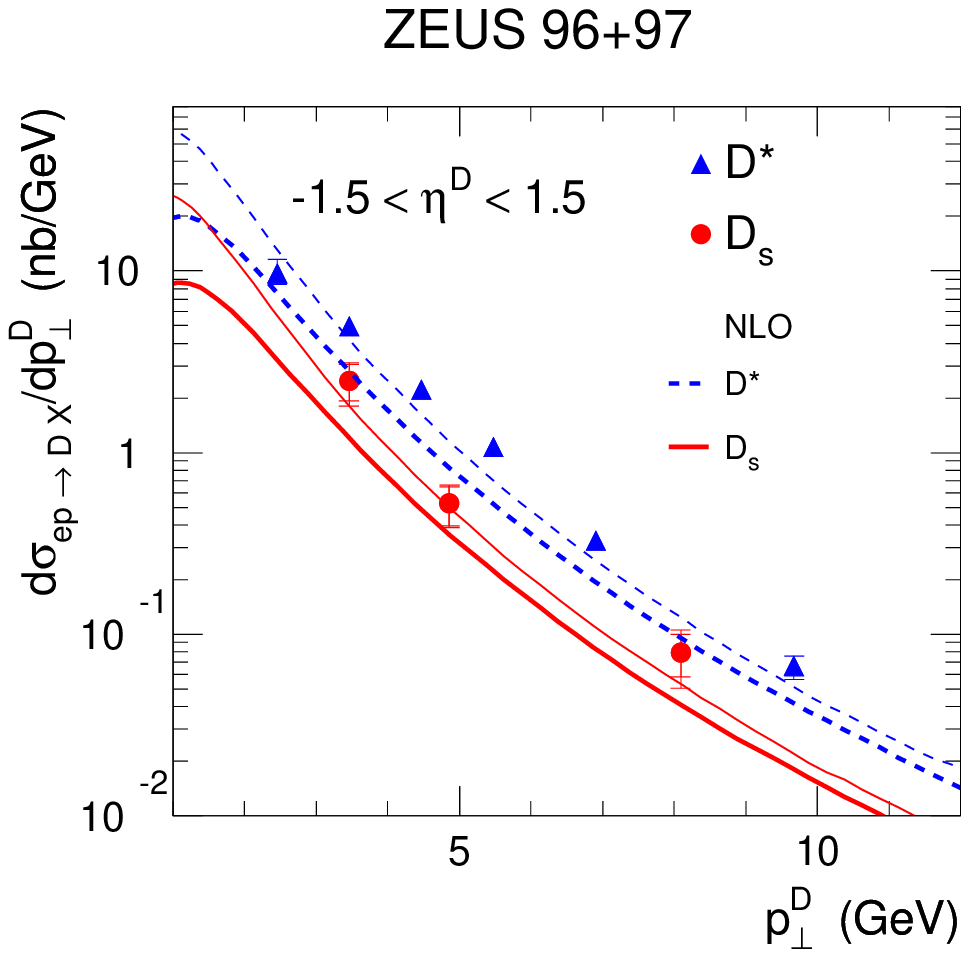,height=5.0cm,width=5.0cm}}
 
\vspace*{-5.0cm}
\hspace*{8.0cm}
{\epsfig{file=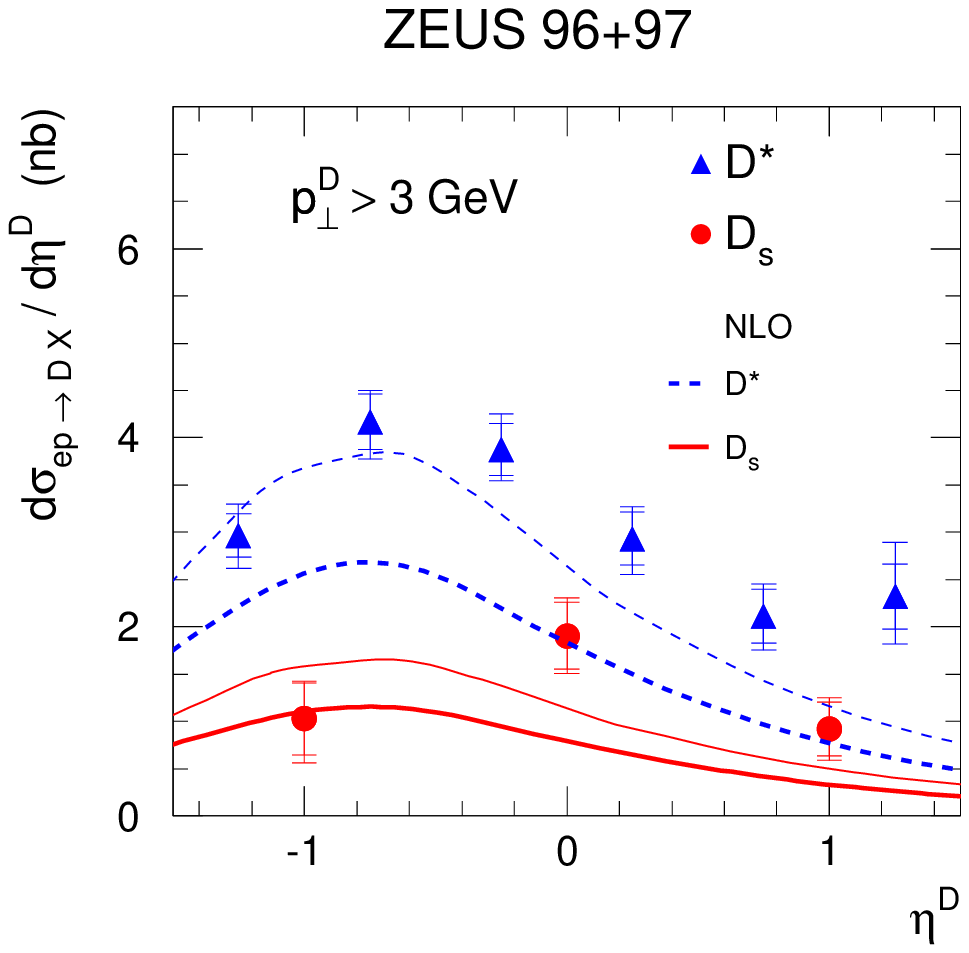,height=5.0cm,width=5.0cm}}
 
\vspace*{-0.1cm}
\bf {~~~~~~~~~~~~~~~~~~~~~~~~~~~~~~~~~(a)~~~~~~~~~~~~~~~~~~~~~~~~~~~~~~~(b)}
\vspace{10pt}   
\caption{Differential cross sections for the photoproduction  
reaction $e p  \to D X$: (a) $d\sigma$/$dp_\perp^D$ and (b)
$d\sigma$/$d\eta^D$, where $D$ stands for  
$\dstar$ or $\dsubs$.  Inner (outer) error bars show statistical (statistical  
and systematic added in quadrature) errors. The $\dsubs$ (dots) and $\dstar$
(triangles) data are compared with NLO predictions for $\dsubs$ (full curves)  
and $\dstar$ (dashed curves) with two parameter settings: $m_c = 1.5 \,
\gev$, $\mu_{R} = m_{\perp}$ (thick curves) and $m_c = 1.2 \, \gev$,  
$\mu_{R} = 0.5 m_{\perp}$ (thin curves).}
\label{fig2}
\end{figure}
The ratio of the cross section for $\dsubspm$ to $\dstarpm$ production at
HERA has been compared to that from $e^{+}e^{-}$ experiments, where the 
latter result is taken from a recent compilation 
\cite{lgcomp}
of fragmentation fractions to charm mesons ($f(c \to D)$).
The results from the two types of interactions are:   \\    
$\xsepdsubs$/$\xsepdstar = \xsecra$   and 
$f(c \to \dsubs^{+})$/$f(c \to \dstarplus) = 0.43 \pm 0.04$. \\
The strangeness suppression factor, $\gamma_s$, (the ratio of the probability
to produce a strange quark to that to produce a non-strange quark), has also
been compared to that from $e^{+}e^{-}$ experiments. 
From HERA the value of the above cross section ratio and
the PYTHIA Monte Carlo was used and for $e^{+}e^{-}$ the quantity
${2f(c \to \dsubs^{+}})$/${[f(c \to \d0)+f(c \to \dplus)]}$
served as an estimator for $\gamma_s$.  The values of $\gamma_s$
for HERA and $e^{+}e^{-}$, respectively, were  
$0.27 \pm 0.04 ^{+0.02}_{-0.03}$ and $0.26 \pm 0.03$,        
implying consistency with universal charm fragmentation.
 
 
\section*{Excited Charm Mesons}
As a basis for the study of higher excitatations of charm mesons, an
enlarged sample of data (integrated luminosity of $110 \picb$), containing
a clean signal of
$\dstarpm$ mesons from both photoproduction and deep inelastic scattering,       
was used  
\cite{osakad}.
Events in the mass range $1.83  <  M(K\pi)   <  
1.90 \, \gev$, $0.144  <  M(K\pi\pi_s) - M(K\pi)  <  0.147 \, \gev$
were chosen  ($\pi_s$ is the low momentum pion in the $\dstar$ decay).
The background (estimated from events in which the K and $\pi$ in the $\d0$ 
mass range have the same charge) has been subtracted, 
yielding $27286 \pm 232$ $\dstarpm$.
 
For orbital excitations an extra track, $\pi_4$, was
added to the $\dstarpm$ candidate and the effective mass combination,    
$\kpiext - \kpipis +  M(\dstar)(2.010 \, \gev)$, was evaluated.
An enhancement in the mass distribution with total charge zero
is seen in Fig. 3a.
This spectrum was fitted to $D_1^0$ and $D_2^{* 0}$
Breit-Wigner shapes with masses and widths fixed 
\cite{pdg},
and convoluted with a Gaussian function with a width as in the Monte Carlo 
simulation.  The background was described by 
$x^\alpha e^{-\beta x + \gamma x^2}$, where
$x = \kpiext - \kpipis - m_{\pi}$, with $\alpha$, $\beta$ and $\gamma$
constant.  Helicity angle distributions for $D_1^0$ and $D_2^{* 0}$ 
proportional to $1+3\cos^2\theta$ and $1-\cos^2\theta$, respectively, 
were folded in for the fit.  
\begin{figure}[h!] 
\unitlength1cm
\hspace*{2.0cm}
{\epsfig{file=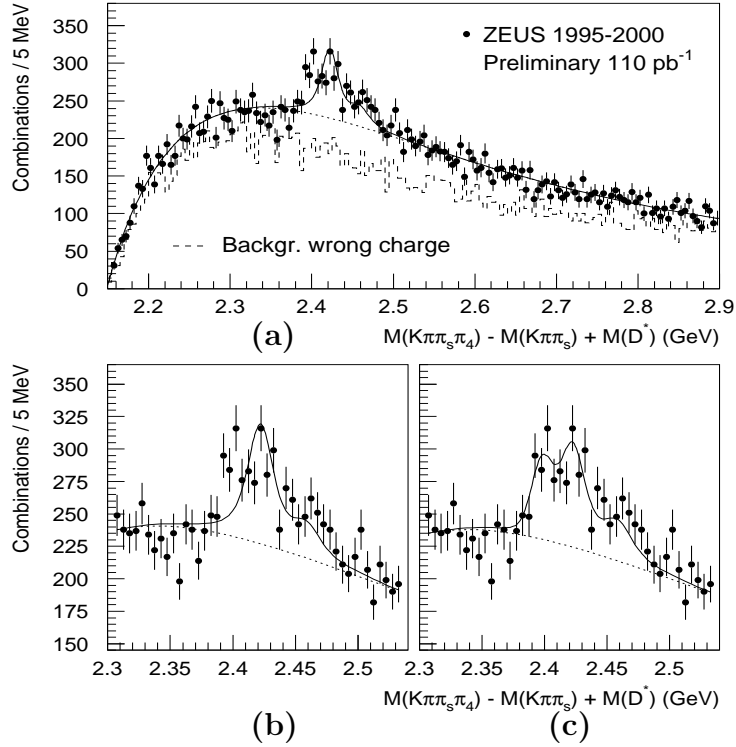,height=11.3cm,width=11.3cm}}
 
\vspace*{-5.9cm}
\bf {~~~~~~~~~~~~~~~~~~~~~~~~~~~~~~~~~~~~(a)}
 
\vspace*{4.7cm}
\bf {~~~~~~~~~~~~~~~~~~~~~~~~~~~~~~~~~~~~(b)~~~~~~~~~~~~~~~~~~~~~(c)}
\vspace{10pt}   
\caption{Distribution of $\kpiext - \kpipis +  M(\dstar)(2.010 \, \gev)$
for $\dstar$ and $\pi_4$ with opposite charges. 
(a)-(b) Fit to two Breit-Wigner shapes convoluted with a Gaussian (full curve). 
The dashed histogram is for $\dstar$ and $\pi_4$ with the same
charge. (c) Fit included extra Gaussian with free mass and width.  The dotted 
curves are the fitted combinatorial background.}
\label{fig3}
\end{figure}
Here, $\theta$ is the angle between $\pi_4$ and $\pi_s$ in the $\dstarpm$ rest
frame.  A closer look, (Fig. 3b), indicates an 
excess of events near $2.4 \, \gev$.  An extra Gaussian  was included to
better fit the data.  The fit (Fig. 3c) yielded
$526 \pm 65$, $203 \pm 60$, and $211 \pm 49$ entries for the number of
$D_1^0$, $D_2^{* 0}$ and the additional Gaussian combinations.  The mass
of the extra Gaussian was 
$2398.1\pm2.1^{+1.6}_{-0.8} \mev$ with a width consistent
with the detector resolution.
 
The ratios of the rates of the $\dstarpm\pi^{\mp}$ decay channel of 
$D_1^0$ and $D_2^{* 0}$ to $\dstarpm$ are
\mbox{$3.40\pm0.42^{+0.78}_{-0.63}\%$} and
\mbox{$1.37\pm0.40^{+0.96}_{-0.33}\%$}, respectively.
Extrapolating to the full kinematic region and using
\cite{lgcomp}
along with known branching ratios
\cite{pdg}
as well as isospin conservation, the result is given in 
table 1.
This result is consistent with those from $e^{+}e^{-}$ experiments.
\begin{table}[h!]
\caption{Comparison of $D_1^0$ and $D_2^{* 0}$ production rates. For ZEUS
errors are statistical, systematic and extrapolation errors.  For CLEO, OPAL
and ALEPH the statistical and systematic errors have been added in quadrature.
The DELPHI results are without systematic errors.}
\label{table1}
\Large{
\begin{tabular}{lccc}         
 Experiment
 & $f(c \to D_1^{0})$\   $[\%]$ & $f(c \to D_2^{* 0})$\ $[\%]$ \\
\hline
ZEUS Prelim.
& $1.46\pm0.18^{+0.33}_{-0.27}\pm0.06$
& $2.00\pm0.58^{+1.40}_{-0.48}\pm0.41$ \\ 
CLEO  \cite{cleo2}
& $1.8\pm0.3$
& $1.9\pm0.3$ \\
OPAL  \cite{opal2}
& $2.1\pm0.8$
& $5.2\pm2.6$ \\
ALEPH Prelim.  \cite{aleph2}
& $1.6\pm0.5$
& $4.7\pm1.0$ \\
DELPHI Prelim.  \cite{delphi2}
& $1.9\pm0.4$
& $4.7\pm1.3$ \\
\hline
\end{tabular}
}
\end{table}
 
In order to search for a radially excited $\dstar$ meson, $D^{*'}$, 
the combination 
$\kpiexte - \kpipis +  M(\dstar)$, 
where $\pi_4$ and $\pi_5$ are oppositely charged pions with 
$p_{\perp} > 0.125 \gev$, was fitted along with a background 
distribution of $x^\alpha e^{-\beta x}$, where
$x = \kpiexte - \kpipis - 2m_{\pi}$.
No peak was seen in the expected mass range,    
($2.59 - 2.67 \, \gev$) (Fig. 4).  An upper limit was obtained by
\begin{figure}[h!] 
\unitlength1cm
\hspace*{3.7cm}
{\epsfig{file=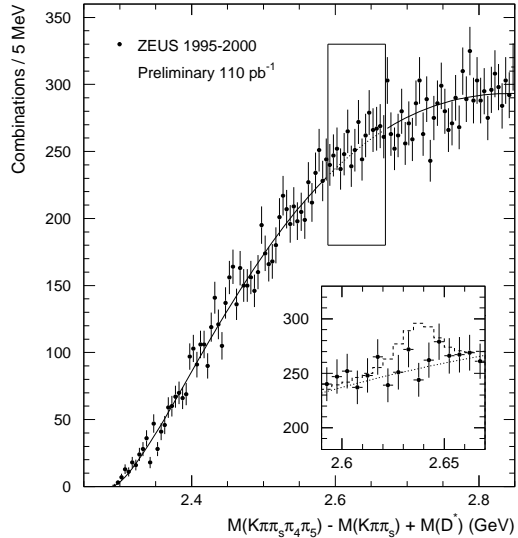,height=8.3cm,width=8.3cm}}
\caption{Distribution of $\kpiexte - \kpipis +  M(\dstar)$ for the
$\dstarpm$ candidates with the $D^{*'}$ window within the rectangle.
Inset:  Dashed histogram is the Monte Carlo
signal normalized to the upper limit and added to the fit
interpolation (dotted curve) in the $D^{*'}$ window.}
\label{fig4}
\end{figure}
fitting the background outside this range, interpolating within the range
and subtracting this from the data in the mass range.  The $95\%$ confidence 
level upper limit for the $\dstarpm\pi^{+}\pi^{-}$ decay relative to 
$\dstarpm$ was found to be $2.3\%$.  Extrapolating to the full
kinematic range and using 
\cite{lgcomp},  
$f(c \to \dstprplus) \cdot B_{\dstprplus \to \dstarplus \pi^{+} \pi^{-}} < 0.7 \%$
at $95\%$ confidence level was obtained.  The equivalent OPAL limit is $1.2\%$ 
\cite{opcleo}.
\section*{Inelastic $\jpsi$ Production}
The $\mu^{+}\mu^{-}$ decay channel of $\jpsi$ for $0.4 < \, z \, < 0.9$, 
$50 \, < \, W_{\gamma p} \, < \, 180 \, \gev$ and $\qsq \, < \, 1 \, \gev^2$, 
using the 1996-1997 data, has been studied 
\cite{osakaj}.
Distributions in $z$, rapidity and transverse momentum squared of the $\jpsi$
and a comparison with theoretical expectations are shown in Fig. 5.
The $z$ dependence of the data is not described in
magnitude by the LO color singlet and octet model with octet matrix
elements calculated from the CDF data
\cite{kniehl,cdf3}.
On the other hand, the NLO color singlet model
\cite{kramer} 
roughly fits the spectrum for $p_{\perp\psi} > 1 \, \gev$.
For $p_{\perp\psi} > 2 \, \gev$ the LO model with octet matrix elements
from CLEO data agrees with the data for high $z$ only 
\cite{beneke,cleo3}.
Currently there is no calculation in NLO for the rapidity distribution.
The NLO calculation agrees with the $p_{\perp\psi}^2$ data.
\begin{figure}[h!] 
\unitlength1cm
{\epsfig{file=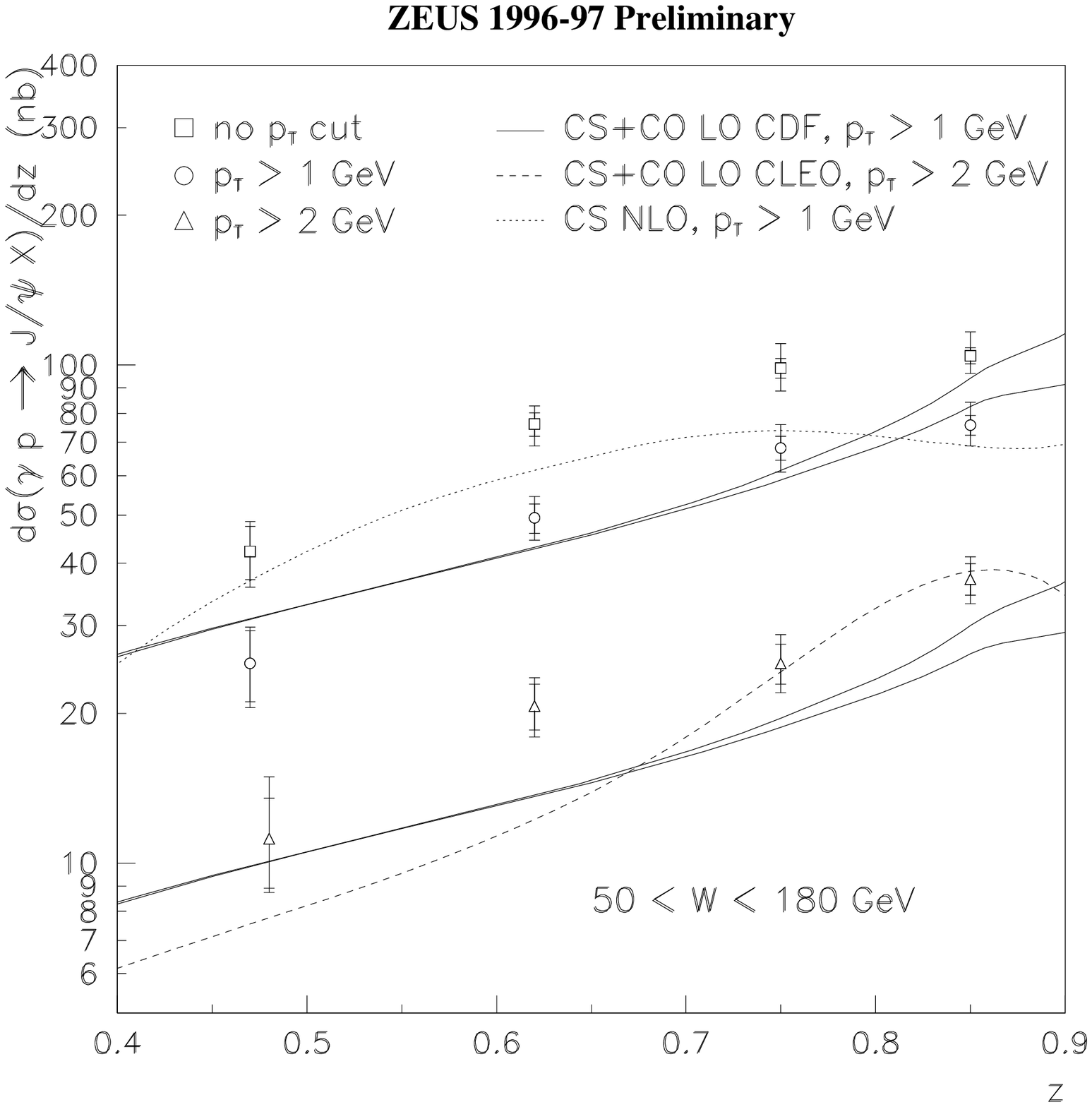,height=6.5cm,width=5.7cm}}
 
\vspace*{-6.5cm}
\hspace*{4.5cm}
{\epsfig{file=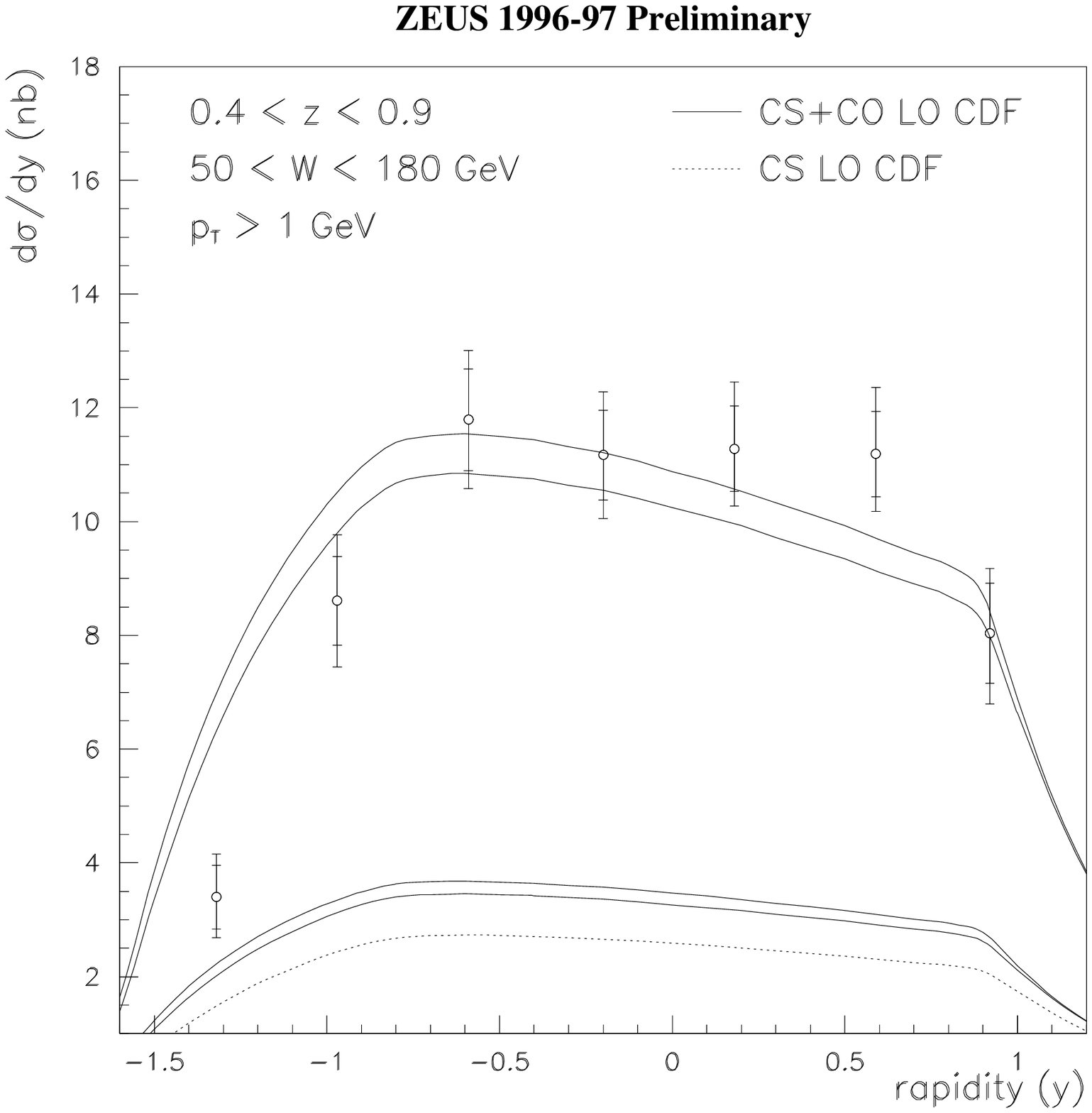,height=6.5cm,width=5.7cm}}
 
\vspace*{-6.5cm}
\hspace*{9.5cm}
{\epsfig{file=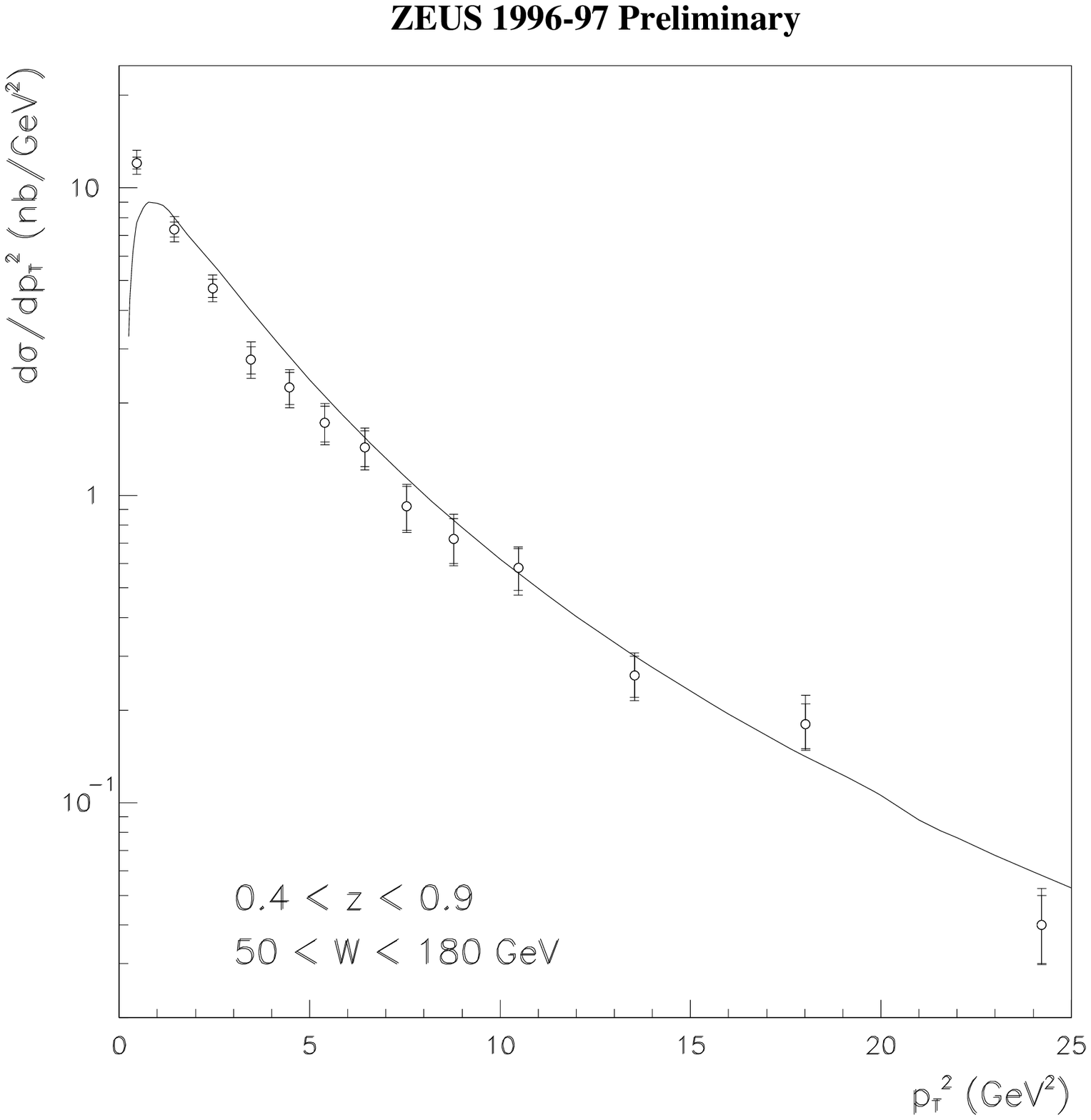,height=6.5cm,width=5.7cm}}
 
\vspace*{-0.7cm}
\bf {~~~~~~~~~~~~~~~ (a) ~~~~~~~~~~~~~~~~~~~~~~~~~(b) ~~~~~~~~~~~~~~~~~~~~~~~~~ (c)}
\vspace{10pt}   
\caption{(a) $z$ distribution for various $p_{\perp\psi}$ cuts:  
no cut (squares), $p_{\perp\psi} > 1 \, \gev$ (circles) and   
$p_{\perp\psi} > 2 \, \gev$ (triangles).  Inner (outer) error bars  
are statistical (quadratic sum of statistical and systematic) errors. 
Lower pair of solid curves are a prediction of color singlet and 
octet model
\protect\cite{kniehl}  
for $p_{\perp\psi} > 1 \, \gev$.  Separation of the curves indicates the  
uncertainty in the color octet matrix elements.  Upper pair of solid 
curves includes a scale factor of $\sim 3$.  Dotted curve is the color   
singlet NLO prediction for the direct photon process and
$p_{\perp\psi} > 1 \, \gev$      
\protect\cite{kramer}.   
Dashed curve is prediction of the
color singlet and octet models for
$p_{\perp\psi} > 2 \, \gev$  
\protect\cite{beneke}.
(b) Rapidity distribution for 
$p_{\perp\psi} > 1 \, \gev$.  Solid curves as in (a).  Dotted curve 
is the LO contribution of the direct photon color singlet component.  
(c) $p_{\perp\psi}^2$ distribution: Solid curve is the prediction of
the NLO calculation 
\protect\cite{kramer}.   }
\label{fig5}
\end{figure}
\section*{Acknowledgment}
It is a pleasure to thank the organizers for a stimulating and
enjoyable meeting.


\begin{references}
\bibitem{detectr}ZEUS Collaboration, Derrick M. et al., {\it Phys. Lett.}
\ {\bf B293}, 465 (1992). \\ 
{\it The ZEUS Detector: Status Report 1993},DESY 1993.
\bibitem{dsubsrf}ZEUS Collaboration, Breitweg J. et al., {\it Phys. Lett.}
\ {\bf B481}, 213 (2000).
\bibitem{dstarrf}ZEUS Collaboration, Breitweg J. et al., {\it Eur. Phys. J.}
\ {\bf C6}, 67 (1999).
\bibitem{pdg}Caso C. et al., Particle Data Group, {\it Eur. Phys. J.}
\ {\bf C3}, 1 (1998).
\bibitem{hqet}Isgur N. and Wise M. B., {\it Phys. Lett.}
\ {\bf B232}, 113 (1989);   
             Neubert N., {\it Phys. Reports.}
\ {\bf A245}, 259 (1994). 
\bibitem{cleo1}CLEO Collaboration, Anderson S. et al., {\it Nucl. Phys.}
\ {\bf A663}, 647 (2000).    
\bibitem{delphi}DELPHI Collaboration, Abreu P. et al., {\it Phys. Lett.}
\ {\bf B426}, 231 (1998).    
\bibitem{opcleo}OPAL Collaboration, submitted to the XXIX  
 International Conference on High Energy Physics, ICHEP 98, 
 Vancouver Canada,July 1998;
\ {\it OPAL PN}, 352.         
\bibitem{rodr}CLEO Collaboration, Rodriquez J. L. {\it hep-ex.}
\ {\bf 9901008}. 
\bibitem{psizeus}ZEUS Collaboration, Breitweg J. et al., {\it Z. Phys.}
\ {\bf C76}, 599 (1997);   
ZEUS Collaboration, contributed paper 814 to the XXIX ICHEP Conference,
Vancouver, Canada (1998).                                      \
\bibitem{psiresolv}Jung H., Schuler A. and Terron J.,
 ,{\it Int. Jour. of Mod. Phys.}
\ {\bf A7}, 7955 (1992).
\bibitem{frix}Frixione S. et al., {\it Nucl. Phys.}
\ {\bf B454}, 3 (1995); {\it Phys. Lett}
\ {\bf B348}, 633 (1995).
\bibitem{lgcomp}Gladilin L., {\it hep-ex.}
\ {\bf 9912064}. 
\bibitem{osakad}ZEUS Collaboration, contributed paper 448 to the XXX ICHEP 
 Conference, Osaka, Japan (2000).
\bibitem{cleo2}CLEO Collaboration, Avery P. et al., {\it Phys. Lett.}
\ {\bf B331}, 236 (1994).
\bibitem{opal2}OPAL Collaboration, Ackerstaff K. et al., {\it Z. Phys.}
\ {\bf C76}, 425 (1997).
\bibitem{aleph2}ALEPH Collaboration,  {\it Production of $D_1$ and $D_2^{*}$
   mesons in hadronic Z decays}, Contributed paper to HEP99, Tampere, 
   July 15-21, Abstract 5\_411.
\bibitem{delphi2}DELPHI Collaboration,  {\it Narrow $D^{**}$ production
   in c and b jets}, Contributed paper to ICHEP98, Vancouver, July 23-29,  
   Paper 240.
\bibitem{osakaj}ZEUS Collaboration, contributed paper 446 to the XXX ICHEP 
 Conference, Osaka, Japan (2000).
\bibitem{kniehl}Kniehl B. A. et al., {\it Eur. Phys. J.}
\ {\bf C6}, 493 (1999);   
Kniehl B. A., {\it Proceedings of the Workshop 1998-1999 on Monte Carlo
Generators for HERA Physics}, ed. by Doyle A. T., Grindhammer G., Ingelman G.  
and Jung H., DESY-PROC-1999-02,p.427.
\bibitem{kramer}Kramer M. et al., {\it Phys. Lett.}
\ {\bf B348}, 657 (1995);  
 Kramer M., {\it Nucl. Phys.}
\ {\bf B459}, 3 (1996).                           
\bibitem{beneke}Beneke M. et al., {\it hep-ex.}
\ {\bf 0001062}, subitted to {\it Phys. Rev. D}.
\bibitem{cdf3}CDF Collaboration, Abe F. et al., {\it Phys. Rev. Lett.}
\ {\bf 79}, 578 (1997).                           
\bibitem{cleo3}CLEO Collaboration, Balest R. et al., {\it Phys. Rev.}
\ {\bf D52}, 2661 (1995).                           
 
\end{references}
\end{document}